\documentclass[a4paper,12pt]{article}

\usepackage{amsmath}
\usepackage{amssymb}
\usepackage{textcomp}
\usepackage{hyperref}
\usepackage{cite}
\title{Quantization of Complementary ADHM Sigma Model}
\author{Abbas Ali, Mohsin Ilahi and Shafeeq Rahman Thottoli  \\
	Physics Department, Aligarh Muslim University,\\ Aligarh, India }
\date{}

\begin{document}

\maketitle
\begin{abstract}
	We discuss quantization of complementary ADHM sigma model in parallel with the quantization of the Witten's original ADHM sigma model by Lambert. The mechanisms of divergence cancellation in two models are different from each other. 
\end{abstract}

\section{Introduction}

Strings can be studied either in flat or curved backgrounds. Latter studies employ the technology of sigma models. These can be linear, that is, massive or non-linear. Alternatively these could be chiral or non-chiral. Chiral sigma models are more difficult to construct and analyze but owing to their phenomenological relevance these deserve to be studied on their own.  Construction of general $(p,q)$ symmetric massive sigma model was done in Ref.\cite{Hull:1993ct,Papadopoulos:1993mf} and their quantization was studied in Ref.\cite{Lambert:1995hs}. A set of interesting backgrounds is provided by the field theoretic instanton constructions. Taking advantage of the earlier construction and quantization of sigma models in  Ref.\cite{Howe:1986vm,Howe:1987qv}  Callan, Harvey and Strominger discussed the string theoretic  generalizations of 't Hooft instanton by solving the relevant heterotic string low energy effective field equations in \cite{Callan:1991dj,Callan:1991ky,Callan:1991at}. The next natural thing was to discuss the massive sigma model with (0,4) supersymmetry corresponding to ADHM instanton \cite{Atiyah:1978ri}. This was done  by Witten in Ref. \cite{Witten:1994tz}. Quantization of the resulting ADHM instanton sigma model was discussed in Ref. \cite{Lambert:1995dp}. A D-brane version was proposed by Douglas in Ref.\cite{Douglas:1996uz} (see also \cite{Johnson:1998yw}). The gauge 1-brane and 5-brane solutions were discussed in \cite{Lambert:1996yd}. In Ref.\cite{Ali:2023csc}  Ali and Ilahi constructed a linear ADHM sigma model that is related to the model constructed by Witten by a simple duality. This model is the one that is being called complementary ADHM sigma model in the title of the present note. Both models, Witten's original model and Ali-Ilahi's complementary model, are linear, that is, massive sigma models. There is a duality between them that  was already perceptible in Ref.\cite{Lambert:1997gs}. In the infrared limit the two models flow to different conformal limits. Ali and Salih constructed a complete  ADHM  linear sigma  model in Ref.\cite{Ali:2023icn} (see also \cite{Ali:2023xov, Ali:2023kkf})according to the suggestions made by Witten in Ref.\cite{Witten:1994tz}. In the present note we shall discuss the quantization of the complementary ADHM linear sigma model  in parallel with Lambert's procedure. In our construction we shall use the duality between the Original and Complementary Models. Though the duality between the two models is simple but its implementation has sufficient novelty to warrant an independent treatment and do the book keeping. 

First of all we shall collect the requisite background and ingredients in Section 2. In Section 3 we discuss the dynamics of ultraviolet and infrared divergences. The task of integrating out the massive modes is taken up in Section 4. The   effect of anomalies is dealt with in Section 5. Finally in Section 6 the case of small instanton is discussed.

\section{Background and Ingredients}
We begin by introducing the action for Witten's original ADHM sigma model.
It has $4k$ bosons $X^{AY}$, $A = 1,2,$ $Y = 1,2,...,2k$ and $4k'$ bosons  $\phi^{A'Y'}$, $A' = 1,2,$ $Y'= 1,2,...,2k'$ and corresponding right-handed superpartners $\psi_{-}^{A'Y}$ for  $X^{AY}$, and $\chi_{-}^{AY'}$ for $\phi^{A'Y'}$. There
are also left-handed fermions $\lambda_{+}^{a}$, $a=1,2,...,n$. The indices  are
raised and lowered by taking the help of antisymmetric tensors $\epsilon^{AB} (\epsilon_{AB})$, $\epsilon^{A'B'} (\epsilon_{A'B'})$, $\epsilon^{YZ} (\epsilon_{YZ})$ and $\epsilon^{Y'Z'} (\epsilon_{Y'Z'})$ belonging to $SU(2)$, $SU(2)'$, $Sp(2k)$ and
$Sp(2k')$ respectively.

  The supersymmetry transformations are
\begin{eqnarray}\label{susytransform}
\delta X^{A Y} & = & i\epsilon_{A'B'} \eta_+^{AA'}\psi_-^{B'Y},   ~~~~ \delta \psi_-^{A'Y}    = \epsilon_{AB} \eta_+^{AA'}\partial_{-}X^{BY} , \cr
~\delta \phi^{A'Y'}  & = &i\epsilon_{AB} \eta_+^{AA'}\chi_-^{BY'}, ~~~~~~  \delta \chi_-^{A'Y}  = \epsilon_{A'B'} \eta_+^{AA'}\partial_{-}\phi^{B'Y'} , \cr
\delta\lambda^a_+ &= &\eta_+^{A A^{\prime}} C_{A A^{\prime}}^a	
\end{eqnarray}
where $\eta_+^{AA'}$ is the infinitesimal anti-commuting parameter. 

The action for the Complementary model is
\begin{equation}\label{action4}
 S_C=\hat S^{kin}+\hat S^{int}
\end{equation}
with 
\begin{eqnarray}\label{action5}
 \hat S^{kin} &=& \int d^2 \sigma ( \epsilon_{AB} \epsilon_{YZ} \partial_{-}  X^{AY}\partial_{+} X^{BZ}+ i\epsilon_{A^{\prime}B^{\prime}}\epsilon_{YZ}\psi_-^{A^{\prime}Y} \partial_{+} \psi_-^{B^{\prime}Z}  \nonumber \\
&+&\epsilon_{A^{\prime}B^{\prime}}\epsilon_{Y^{\prime}Z^{\prime}}\partial_{-}\phi^{A^{\prime}B^{\prime}}\partial_{+} \phi^{B^{\prime}Z^{\prime}}
+i\epsilon_{AB}\epsilon_{Y^{\prime}Z^{\prime}}\chi_-^{AY^{\prime}}\partial_{+} \chi_{-}^{BZ^{\prime}} \nonumber \\&+&i\hat \lambda_+^{a^\prime} \partial_{-}\hat \lambda_+^{a^\prime})
\end{eqnarray}
and
\begin{eqnarray}\label{action6}
    \hat S^{int}&=&-\frac{i}{2}m\int d^2 \sigma\hat\lambda^{a^{\prime}}_{+}[(\epsilon^{BD} \frac{\partial \hat C^{a^{\prime}}_{BB'}}{\partial X^{DY}}\psi^{B'Y}_{-} + \epsilon^{B'D'}\frac{\partial \hat C^{a'}_{BB'}}{\partial \phi^{D'Y'}}{\chi}^{BY'}_{-})\nonumber \\&-& \frac{im}{4} \epsilon^{AB}\epsilon^{A'B'}\hat C^{a'}_{AA'}\hat C^{a'}_{BB'})]
\end{eqnarray}

with the conditions
\begin{eqnarray}\label{adhm3}
\frac{\partial \hat C_{AA'}^{a'}}{\partial X^{B Y}}
+\frac{\partial \hat C_{BA'}^{a'}}{\partial X^{A Y}}
=0=  \frac{\partial \hat C_{AA'}^{a'}}{\partial \phi^{B'Y'}}
+\frac{\partial \hat C_{AB'}^{a'}}{\partial \phi^{A'Y'}},
\end{eqnarray}

\begin{equation}\label{adhm4}
\sum_{a'}(\hat C^{a'}_{AA'}{\hat C}^{a'}_{BB'}+{\hat C}^{a'}_{BA'}{\hat C}^{a'}_{AB'})=0
\end{equation}
with $\hat C^{a'}_{AA'}$ given by
\begin{equation}\label{generalform2}
\hat C^{a'}_{AA'}=\hat M^{a'}_{AA'}+\epsilon_{AB} \hat N^{a'}_{A'Y}X^{BY}+\epsilon_{A'B'}\hat D^{a'}_{AY'}\phi^{~B'Y'}+\epsilon_{AB}\epsilon_{A'B'}\hat E^{a'}_{YY'}X^{BY}\phi^{~B'Y'}.
\end{equation}

For Witten's original ADHM sigma model the choice is
$M^{a}_{AA'}=N^{a}_{A'Y}=0$ such that the tensor $C^{a}_{AA'}$ reduces to
\begin{equation}\label{caaa}
C^{a}_{AA'}=\epsilon_{A'B'}(D^{a}_{AY'}+\epsilon_{AB}E^{a}_{YY'}X^{BY})\phi^{B'Y'}=\epsilon_{A'B'}B^{a}_{AY'}(X)\phi^{B'Y'}
\end{equation}
where  $B^{a}_{AY'}(X)$ is linear in $X$ because the tensors  $D^{a}_{A'Y}$ and $E^{a}_{YY'}$ are constant. For the Complementary model we have $\hat M^{a'}_{AA'}=\hat D^{a'}_{AY'}=0$ and the tensor $\hat C^{a'}_{AA'}$ reduces to
\begin{equation}\label{cnex}
\hat C^{a'}_{AA'}=\epsilon_{AB}(\hat N^{a'}_{A'Y}+\epsilon_{A'B'}\hat E^{a'}_{YY'}\phi^{B'Y'})X^{BY}=\epsilon_{AB}A^{a'}_{A'Y}(\phi)X^{BY}.
\end{equation}
This time $A^{a'}_{A'Y}(\phi)$ is linear in $\phi$ because $\hat N^{a'}_{A'Y}$ and $\hat E^{a'}_{YY'}$ are constant. 

Both of the models have $(0,4)$ supersymmetry. In Ref. \cite{Lambert:1995dp} a $(0,1)$ superfield formalism was proposed for the original model on the basis of the formalism of \cite{Hull:1993ct,Papadopoulos:1993mf,Galperin:1994qn}. It is valid in the complementary case too. We introduce the complex structure $J^{AA'}$ with properties $\epsilon_{AB}J^{A}_{A'}J^{B}_{B'}=\epsilon_{A'B'}$, $J^{AB'}J_{AC'}=-\delta^{B'}_{C'}$ and $J^{BA'}J_{CA'}=-\delta^{B}_{C}$. The superfields in the original case are
\begin{eqnarray}\label{superfields1}
\mathcal{X}^{AY} &=& X^{AY} + \theta^-J^A_{\ A'}\psi_-^{A'Y},~	\Phi^{A'Y'} = \phi^{A'Y'} + \theta^-J^{\ A'}_{A}\chi_-^{AY'}\nonumber\\
\Lambda^a_+ &=& \lambda^{a}_{+} + \theta^{-}F^a.
\end{eqnarray}
In the Complementary case the third field above is replaced by 
\begin{equation}\label{superfields2}
    \hat \Lambda^{a^{\prime}}_{+} = \hat \lambda^{a^{\prime}}_{+} + \theta^-\hat F^{a^{\prime}}.
	\end{equation}
The superderivative is $
D_- = \frac{\partial}{\partial \theta^-} + i\theta^-\partial_{-}$ in both cases. The auxiliary fields $F^a$ and $\hat F^{a'}$  are removed using their respective equations of motion. The actions in superspace become
\begin{eqnarray}\label{seff1}
S_{eff} &=& -i\int d^2x d\theta^{-}\Bigl(\epsilon_{AB}\epsilon_{YZ}D_{-}\mathcal{X}^{AY}\partial_{+}\mathcal{X}^{BZ}%\nonumber\\&
+\epsilon_{A'B'}\epsilon_{Y'Z'}D_{-}\Phi^{A'Y'}\partial_{+}\Phi^{B'Z'} \nonumber\\&-& i\Lambda^a_{+}D_{-}\Lambda^a_{+}-mC_a\Lambda^a_+\Bigl)
\end{eqnarray}
and
\begin{eqnarray}\label{seff2}
\hat S_{eff} &=& -i\int d^2x d\theta^{-}\Bigl(\epsilon_{AB}\epsilon_{YZ}D_{-}\mathcal{X}^{AY}\partial_{+}\mathcal{X}^{BZ}%\nonumber\\&
+\epsilon_{A'B'}\epsilon_{Y'Z'}D_{-}\Phi^{A'Y'}\partial_{+}\Phi^{B'Z'} \nonumber\\&-& i \hat \Lambda^{a^{\prime}}_{+}D_{-}\hat\Lambda^{a'}_{+}-m\hat C_{a'}\Lambda^{a^{\prime}}_+\Bigl)
\end{eqnarray}
respectively with $C^{a}=J^{AA'} C^{a}_{AA'}$ and $\hat C^{a'}=J^{AA'}\hat C^{a}_{AA'}$.  

The target space in both cases is $R^{4(k+k')}$ and only $(0,1)$ part of the full $(0,4)$ superalgebra closes off-shell. The off-shell closer of the full (0,4) superalgebra requires harmonic superspace formalism. For the original model this was developed in Ref. \cite{Galperin:1995pq} while for the complementary model it is discussed in Ref.\cite{Ali:2025ntc, Ali:2025jcu}.

The resulting instanton for complementary models has the expression
\begin{equation}\label{aijay}
\hat A_{i'j'A'Y'}=\sum_{a'=1}^{n'=N'-4k}\hat v^{a'}_{i'}\frac{\partial \hat v^{ a'}_{j'}}{\partial \phi^{A'Y'}}.
\end{equation}
It should be compared with the corresponding expression for the instanton for the original model 
\begin{equation}\label{aijay0}
 A_{ijAY}=\sum_{a=1}^{n=N-4k'}v^{a}_{i}\frac{\partial v^{a}_{j}}{\partial X^{AY}}.
\end{equation}

In the $SU(2)$ case the potentials in the original and complementary model are 
\begin{equation}\label{potential}
V_W = \frac{m^2}{8}(X^{2}+\rho^{2})\phi^{2}~\text{and}~V_C = \frac{m^2}{8}(\phi^{2}+\omega^{2})X^{2}
\end{equation}
respectively. 

There is a duality between the original and complementary ADHM sigma models with respect to the interchanges $k  \leftrightarrow k', ~~~A  \leftrightarrow A',~~~Y  \leftrightarrow Y',~~~X  \leftrightarrow \phi~~\text{and}~~ \rho \leftrightarrow \omega$. This is most apparent in these expressions.

\section{Divergence Management}

To discuss the quantization of this model we begin with a discussion of ultraviolet divergences. Like  Witten's original ADHM sigma model Ali-Ilahi's complementary model too is  superrenormalizable in two dimensions because the vertices do not carry any momentum factors and have at most three legs. The only possible ultraviolent divergences come from the one-loop graphs contributing to the potential. The superspace argument of Ref. \cite{Lambert:1995dp} is valid for the complementary model too. The measure of superspace $d^{2}xd\theta^{-}$ possesses dimension of mass $-\frac{3}{2}$ whereas every vertex contributes to the effective action a factor of $m$.  By power counting it is clear that only graphs with a single vertex can yield divergent contribution to the effective action. Among all of these only the one loop (tadpole) graphs are relevant and the rest of all the higher loop divergences are removed by the renormalization procedure.

Since massless fields are present in the theory we must pay attention to the infrared divergences too. In order to deal with these divergences an IR regulator is added in the form of a mass $M$ to the propagator and any mass term in the action is treated as interaction and the limit $M\rightarrow 0$ is taken in the final expressions. The graphs are calculated using dimensional regularization in $D=2+\epsilon$ dimensions and the background field method \cite{Howe:1986vm}.

The fermionic and bosonic divergent contributions for the Complementary Model are 
\begin{eqnarray}\label{gdivboson2}
\hat\Gamma_{Div}(b) &=&- \frac{m^2}{4}\Delta(0) [\epsilon^{AB}\epsilon^{CD}\epsilon^{C'D'}\epsilon^{YZ} \frac{\partial \hat C^{a'}_{CC'}}{\partial X^{AY}}\frac{\partial \hat C^{a'}_{DD'}}{\partial X^{BZ}} \nonumber \\&+& \epsilon^{A'B'}\epsilon^{C'D'}\epsilon^{CD}\epsilon^{Y'Z'} \frac{\partial \hat C^{a'}_{CC'}}{\partial \phi^{A'Y'}}\frac{\partial \hat C^{a'}_{DD'}}{\partial \phi^{B'Z'}}]
\end{eqnarray}
and
\begin{eqnarray}\label{gdivfermion2}
\hat\Gamma_{Div}(f) &=& \frac{m^2}{8}\Delta(0) [\epsilon^{AC}\epsilon^{BD}\epsilon^{C'D'}\epsilon^{YZ} \frac{\partial \hat C^{a'}_{CC'}}{\partial X^{AY}}\frac{\partial \hat C^{a'}_{DD'}}{\partial X^{BZ}} \nonumber \\&+& \epsilon^{A'C'}\epsilon^{B'D'}\epsilon^{CD}\epsilon^{Y'Z'} \frac{\partial \hat C^{a'}_{CC'}}{\partial \phi^{A'Y'}}\frac{\partial \hat C^{a'}_{DD'}}{\partial \phi^{B'Z'}}].
\end{eqnarray}

At this point, all tensor expressions involving background fields and the bosonic propagator are computed at zero momentum and the renormalization scale $\mu$ such that

\begin{equation}\label{bosonicprop}
\Delta(0)=-\frac{1}{2\pi\epsilon}-\frac{M^2}{\mu^2} +\text{finite}.
\end{equation}

In this case, using Eqn.(\ref{cnex}) for $\hat C^{a'}_{AA'}$, we get

\begin{eqnarray}\label{partder3}
\frac{ \partial \hat C^{a'}_{CC'}}{\partial X^{AY}}=-\epsilon_{AC}A^{a'}_{C'Y}(\phi),~~~\frac{ \partial \hat C^{a'}_{DD'}}{\partial X^{BZ}}=-\epsilon_{BD}A^{a'}_{D'Z}(\phi),
 \end{eqnarray}
 and
 \begin{eqnarray}\label{partder4}
 \frac{\partial \hat C^{a'}_{CC'}}{\partial \phi^{A'Y'}}=\epsilon_{BC}\epsilon_{A'C'}\hat E^{a'}_{YY'}X^{BY},~~~\frac{\partial \hat C^{a'}_{DD'}}{\partial \phi^{B'Z'}}=\epsilon_{BD}\epsilon_{B'D'}\hat E^{a'}_{YZ'}X^{BY}.
 \end{eqnarray}
Thus the first term in $\hat \Gamma_{Div}(b)$ for the Complementary model becomes

\begin{equation}\label{cfbt}
=-\frac{m^2}{4}\hat{b}_1\Delta(0)\epsilon^{C'D'}\epsilon^{YZ}A^{a'}_{C'Y}(\phi)A^{a'}_{D'Z}(\phi)
\end{equation}
%.......................MAKE ARRAY LATER.........................
with  $\hat b_1=(\epsilon^{AB}\epsilon_{AC})(\epsilon^{CD}\epsilon_{BD})$. The corresponding term in $\hat \Gamma_{Div}(f)$ is

\begin{equation}\label{cfft}
=\frac{m^2}{8}\hat f_1\Delta(0)\epsilon^{C'D'}\epsilon^{YZ}A^{a'}_{C'Y}(\phi)A^{a'}_{D'Z}(\phi)
\end{equation}
%.......................MAKE ARRAY LATER.........................
with $\hat f_1=(\epsilon^{AC}\epsilon_{AC})(\epsilon^{BD}\epsilon_{BD})$. Since $\hat f_1=2\hat b_1$ the contributions from (\ref{cfbt}) and (\ref{cfft}) cancel. Next we take up the second terms. Only the $\epsilon$ terms differ from each other in these and the rest includes the same factor. The second term in $\hat \Gamma_{Div}(b)$ is
\begin{equation}\label{cbst}
=\frac{m^2}{4}\hat b_2\Delta(0)\epsilon_{BC}\epsilon^{Y'Z'}E^{a}_{YY'}E^{a}_{ZZ'}X^{BY} X^{CZ}
\end{equation}
with $\hat b_2=(\epsilon^{A'B'}\epsilon_{A'C'})(\epsilon^{C'D'}\epsilon_{B'D'})$ and second term in $\hat \Gamma_{Div}(f)$ is
\begin{equation}\label{cfst}
=-\frac{m^2}{8}\hat f_2 \Delta(0)\epsilon_{BC}\epsilon^{Y'Z'}E^{a}_{YY'}E^{a}_{ZZ'}X^{BY}X^{CZ}
\end{equation}
%.......................MAKE ARRAY LATER.........................
with  $\hat f_2=(\epsilon^{A'C'}\epsilon_{A'C'})(\epsilon^{B'D'}\epsilon_{B'D'})$. Since $\hat f_2=2\hat b_2$ the contributions from (\ref{cbst}) and (\ref{cfst}) cancel. 

\section{Dealing with Massive Modes}

We now take up the task of integrating out the massive modes. This can be dealt with in two different ways - either by using duality on Lambert's analysis or taking up the complementary branch and follow it in parallel with the original branch. We shall use the latter approach. To avoid cluttering of expressions we shall avoid mentioning the issues related to the quantization of the original branch in this section.

If the number $n'$ of the left handed fermions $\hat \lambda^{a'}$ is bigger than the number $4k$ of the components $\psi_{-}^{A'Y}$ then generically all of $\psi_-$'s get mass. In this case, by supersymmetry,  all of the $X$'s too are massive and $N'=n'-4k$ components of $\hat \lambda^{a'}$ are massless. The vacuum in this case is given by $X^{AY}=0$. 

The definition and orthogonality of the zero modes for the Complementary branch, $\hat v_{i'}^{a'}, i'=1,2,...,N'=n'-4k$
are given by

%If we consider the zero modes $v^{a}_{i}(\phi), i = 1,2 ,..., n - 4k$ of the
%fermion mass matrix, described such that
\begin{equation}\label{corthocond}
\hat v^{a'}_{i}A^{a'}_{A'Y} = 0~~\text{and}~~~~~\hat v^{a'}_{i'}\hat v^{a'}_{j'}=\delta_{i'j'}.
\end{equation}
The massive modes $\hat u^{a'}_{I'}(\phi), I' = 1,2 ,..., 4k$ obey
\begin{equation}\label{cmassivemodes}
\hat u^{a'}_{I'}\hat u^{a'}_{J'} =\delta_{I'J'}~~~\text{and}~~~~ \hat u^{a'}_{I'}\hat v^{a'}_{j'}=0.
\end{equation}
The $\hat \lambda^{a'}_{+}$'s are broken as 
\begin{equation}\label{lamdabroken}
\hat \lambda^{a'}_{+}=\hat v_{i'}^{a'}\hat \zeta_{+}^{i'}+\hat u_{I'}^{a'}\hat \zeta_{+}^{I'}
\end{equation}
such that the $\hat \zeta_{+}^{i'}$ are massless and $\hat \zeta_{+}^{I'}$ massive. We can now re-express the action in Eq.(\ref{action4}) in
two parts
\begin{equation}\label{actionparts}
\hat S=\hat S_{0}+\hat S_{m}, 
\end{equation}
where
\begin{eqnarray}\label{actionmassless}
\hat S_{0}&=& \int d^2 x \lbrace \epsilon_{A'B'} \epsilon_{Y'Z'} \partial_{-}  \phi^{A'Y'}\partial_{+} \phi^{B'Z'}+ i\epsilon_{A^{}B^{}}\epsilon_{Y'Z'}\chi_-^{AY'} \partial_{+} \chi_-^{BZ'}  \nonumber \\
&+& i\hat \zeta_{+}^{i'}(\delta_{i'j'}\partial_{-}\hat \zeta_{+}^{j'}+\hat A_{i'j'A'Y'}\partial_{-}  \phi^{A'Y'}\hat\zeta_{+}^{j'})\rbrace,
\end{eqnarray}
with $\hat A_{i'j'A'Y'}$, defined in (\ref{aijay}),
is the induced $SO(n'-4k)$ connection and 
\begin{eqnarray}\label{actionmassive}
\hat S_{m} &=& \int d^2 x \lbrace \epsilon_{AB} \epsilon_{YZ} \partial_{-}  X^{AY}\partial_{+} X^{BZ}+ i\epsilon_{A'B'}\epsilon_{YZ}\psi_-^{A'Y} \partial_{+} \psi_-^{B'Z}   \nonumber \\
&+& i\hat \zeta_{+}^{I'}\partial_{-}\hat \zeta_{+}^{I'}+i\hat A_{I'J'A'Y'}\partial_{-}  \phi^{A'Y'}\hat \zeta_{+}^{I'}\hat\zeta_{+}^{J'}+2i\hat A_{i'J'A'Y'}\partial_{-}  \phi^{A'Y'}\hat\zeta_{+}^{i'}\hat\zeta_{+}^{J'}                       \nonumber \\
&-& im\epsilon_{AB}\hat v_{i'}^{ a'}\hat E_{YY'}^{a'}\hat \zeta_{+}^{i'}X^{AY}\chi_{-}^{BY'}-im\epsilon_{AB}\hat u_{I'}^{a'}\hat E_{YY'}^{a'}\hat\zeta_{+}^{I'}X^{AY}\chi_{-}^{BY'} \nonumber \\
&-& im\hat u_{I'}^{a'}A_{A'Y}^{a'}\hat \zeta_{+}^{I'}\psi_{-}^{A'Y} -\frac{m^2}{8}\epsilon_{AB}\epsilon^{A'B'}A_{A'Y}^{a'}A_{B'Z}^{a'}X^{AY}X^{BZ}\rbrace
\end{eqnarray}
with
\begin{equation}\label{caijay}
\hat A_{I'J'A'Y'}=\hat u^{a'}_{I'}\frac{\partial \hat u^{a'}_{J'}}{\partial \phi^{A'Y'}} ~~~\text{and} ~~~~
 \hat A_{i'J'A'Y'}=\hat v^{a'}_{i'}\frac{\partial \hat u^{a'}_{J'}}{\partial \phi^{A'Y'}}.
\end{equation}

The classical low-energy effective action is simply achieved by taking the most
general action possible that agree with every symmetry that the theory has. 
%To
%compute the effective action quantum mechanically we shall be integrating over the massive
%fields and ignore any higher-derivative terms. The existence of higher-derivative terms
%in the effective action, that are repressed by powers \textit{p/m} where $p$ is the low-energy
%momentum scale, would destroy the renormalizability and avert an easy geometrical
%sigma model interpretation of the effective theory.
To calculate the quantum mechanical effective action, we will perform integration over the massive fields and neglect any terms involving higher derivatives. The presence of higher-derivative terms in the effective action, which are suppressed by factors of $\frac{p}{m}$ where $p$ represents the low-energy momentum scale, would compromise its renormalizability and hinder a straightforward geometric sigma model interpretation of the effective theory.

%Initially we see that due to the non-trivial definition of the massless left-handed
%fermions Eq.(\ref{lamdabroken})
At first, we observe that because of the complex interpretation of massless left-handed fermions indicated in Eq. (\ref{lamdabroken}), $S_{0}$ does not  have $(0,4)$ supersymmetry in itself as it is not having a four-fermion
interaction term. The issue is solved by noticing that there exists a tree graph,
having a single internal $X^{AY}$ field propagating, that adds to the low-energy
effective action. To get rid of the singular behaviour of the propagator at zero
momentum, while computing this graph it is useful to employ the massive propagator for
$X^{AY}$, achieved from the last term in Eq.(\ref{actionmassive}).

Utilising Eq.(\ref{adhm4}) we now express  
\begin{equation}\label{aaomega}
A_{A'Y}^{a}A_{B'Z}^{a}=\epsilon_{A'B'}\epsilon_{YZ}\Omega(\phi) ~~\text{with}~~
\Omega(\phi)=\frac{1}{4k}\epsilon^{A'B'}\epsilon^{YZ}A_{A'Y}^{a}A_{B'Z}^{a}.
\end{equation}
The final term in Eq.(\ref{actionmassive}) becomes $
-\frac{m^2}{4}\Omega(\phi)X^{2}$
and thus could be taken as the $\phi^{A'Y'}$ dependent mass term for $X^{AY}$. The tree graph
could then be noticed to add to the four-fermion term 
\begin{equation}\label{fourferterm}
-\frac{1}{2}\hat \zeta_{+}^{i'}\hat \zeta_{+}^{j'}\hat F_{AY'BZ'}^{i'j'}\chi_{-}^{AY'}\chi_{-}^{BZ'}
\end{equation}
where
\begin{equation}\label{fieldstrength}
\hat F_{AY'BZ'}^{i'j'}=2\epsilon_{AB}\epsilon^{YZ}\Omega^{-1}\hat v_{i'}^{a'}\hat E_{Y(Y'|}^{a'}\hat v_{j'}^{b'}\hat E_{Z|Z')}^{b'},
\end{equation}
that we shall subsequently connect to the field strength tensor. 

We could now ignore all vertices having only one massive field in Eq.(\ref{actionmassive}) and analyze
the one-loop additions to the effective action. Examination of the quadratic terms in $
S_{m}$ reflect that there are no additions to the gauge connection in Eq.(\ref{actionmassless}). Moreover,
examination reflects that of every other possible additions only those corresponding
to the effective potential do not require higher-order derivatives of the massless fields.
A check on this is to notice that any terms that are second order in the derivatives are
logarithmically divergent, and by finiteness of the model, must thus disappear.

To compute the effective potential we just use $\partial_{-} \phi^{A'Y'}=\partial_{+} \phi^{A'Y'}=\chi_{-}^{AY'}=0$. Therefore
only the last two terms in Eq.(\ref{actionmassive}) have to be considered (we no longer utilize a massive
propagator for $X^{AY}$). The effective potential then gets the standard bosonic as well as
fermionic additions (in Euclidean momentum space)
\begin{equation}\label{veffb}
\hat{V}_{eff}(b)=\frac{\alpha'}{4\pi}\sum_{n=1}^{\infty}\frac{1}{n}\int d^{2}p\text{Tr}\left[\frac{\epsilon_{CD}\epsilon^{A'B'}A_{A'Y}^{a}A_{B'Z}^{a}}{4p^2/m^2}\right]^n,
\end{equation}
\begin{equation}\label{vefff}
\hat{V}_{eff}(f)=-\frac{\alpha'}{4\pi}\sum_{n=1}^{\infty}\frac{1}{n}\int d^{2}p\text{Tr}\left[\frac{\hat u_{I'}^{a'}\hat u^{b'I'}A_{C'Y}^{a'}A_{D'Z}^{a'}}{2p^2/m^2}\right]^n.
\end{equation}
Now the definition Eq.(\ref{aaomega}) gives the following equations: 
\begin{equation}\label{aaomega1}
\epsilon_{CD}\epsilon^{A'B'}A_{A'Y}^{a'}A_{B'Z}^{a'}=2\epsilon_{CD}\epsilon_{YZ}\Omega(\phi)
\end{equation}
\begin{equation}\label{aaomega2}
\hat u_{I'}^{a'}\hat u^{b'I'}A_{C'Y}^{a'}A_{D'Z}^{b'}=\epsilon_{C'D'}\epsilon_{YZ}\Omega(\phi).
\end{equation}
Thus Eq.(\ref{veffb}) and Eq.(\ref{vefff}) leaving zero contribution to the
effective potential.

From the above investigation we deduce that the effective quantum action of the massless
fields in the Complementary case is
\begin{eqnarray}\label{effquantact}
\hat S_{eff} &=& \int d^2 x \lbrace \epsilon_{A'B'} \epsilon_{Y'Z'} \partial_{-}  \phi^{A'Y'}\partial_{+} \phi^{B'Z'}+ i\epsilon_{AB}\epsilon_{Y'Z'}\chi_-^{AY'} \partial_{+} \chi_-^{BZ'}  \nonumber \\
&+& i\hat \zeta_{+}^{i'}(\delta_{i'j'}\partial_{-}\hat \zeta_{+}^{j'}+\hat A_{i'j'A'Y'}\partial_{-}  \phi^{A'Y'}\hat \zeta_{+}^{j'})\nonumber\\&-&\frac{1}{2}\hat \zeta_{+}^{i'}\zeta_{+}^{j'}\hat F_{AY'BZ'}^{i'j'}\chi_{-}^{AY'}\chi_{-}^{BZ'}\rbrace.
\end{eqnarray}

%This is just the action of the general $(0,4)$ supersymmetric non-linear sigma model \cite{Hull:1993ct,Papadopoulos:1993mf}
This corresponds to the behavior described by the general $(0,4)$ supersymmetric non-linear sigma model as outlined in \cite{Hull:1993ct,Papadopoulos:1993mf}
, even though the right-handed superpartners of $\phi^{A'Y'}$ are `twisted'. As with the original
theory Eq.(\ref{action4}), the low-energy effective theory Eq.(\ref{effquantact}) allows a $(0, 1)$ superfield form.
Considering the superfield $\hat \Lambda^{i'}_{+} = \hat \lambda_{+}^{i'} + \theta^{-}\hat F^{i'}$ then permits us (after getting rid of $\hat F^{i'}$ by its
equation of motion) to write Eq.(\ref{effquantact}) as
\begin{eqnarray}\label{supereffact}
\hat S_{eff}&=& -i\int d^{2}x\theta^-\lbrace \epsilon_{A'B'}\epsilon_{Y'Z'}D_{-}\Phi^{A'Y'}\partial_{+}\Phi^{B'Z'}\nonumber \\&-&i\hat \Lambda_{+}^{i'}(\delta_{i'j'}D_{-}\hat \Lambda_{+}^{j'}+\hat A_{i'j'A'Y'}D_{-}\Phi^{A'Y'}\hat \Lambda_{+}^{j'}) \rbrace
\end{eqnarray}

if 
\begin{equation}\label{curvature}
\hat F_{AY'BZ'}^{i'j'}=\hat F_{A'Y'B'Z'}^{i'j'} \hat J_{~A}^{A'} \hat J_{~B}^{B'}
\end{equation}
here $\hat F_{A'Y'B'Z'}^{i'j'}$ is the curvature of the connection Eq.(\ref{caijay}),
\begin{equation}\label{faijay}
\hat F_{A'Y'B'Z'}^{i'j'}=\partial_{A'Y'}\hat A_{i'j'B'Z'}-\partial_{B'Z'}\hat A_{i'j'A'Y'}+\hat A_{i'k'A'Y'}\hat A_{k'j'B'Z'}-\hat A_{i'k'B'Z'}\hat A_{k'j'A'Y'}.
\end{equation}
This is simply the known condition on $(0,4)$ models that the field strength be agreeable
with the complex structure \cite{Hull:1993ct,Papadopoulos:1993mf}. Moreover it is not hard to find that $\hat S_{eff}$ does
have the complete on-shell $(0,4)$ supersymmetry (the superspace formulation Eq.(\ref{supereffact}) makes
sure only off-shell $(0,1)$ supersymmetry) exactly when Eq.(\ref{curvature}) is obeyed.

For the $k = k'=1, n = 8$ model the non-zero
components of $\hat v_{i}^{a}$ and $\hat u_{I}^{a}$ are
\begin{eqnarray}\label{nonzerocomp}
\hat v^{YY'}_{ZZ'} = \frac{\omega}{\sqrt{\omega^2 + \phi^2}}\delta_{Z}^{Y}\delta_{Z'}^{Y'}, && \hat v^{A'Y}_{ZZ'} =- \frac{\sqrt2}{\sqrt{\omega^2 + \phi^2}}\phi_{~Z'}^{A'}\delta_{Z}^{Y},\nonumber \\
\hat u^{YY'}_{B'Z} =- \frac{\sqrt2}{\sqrt{\omega^2 + \phi^2}}\phi_{B'}^{~Y'}\delta_{Z}^{Y}, && \hat u^{A'Y}_{B'Z} = \frac{\omega}{\sqrt{\omega^2 + \phi^2}}\delta_{B'}^{A'}\delta_{Z}^{Y}
\end{eqnarray}
whereas the mass term Eq.(\ref{aaomega}) becomes
\begin{equation}
\Omega(\phi)= \frac{1}{2}(\phi^{2}+\omega^{2}).
\end{equation}
The gauge field $A_{ijA'Y'}$ achieved from Eq.(\ref{nonzerocomp}) is just that of a single instanton on the
manifold $R^4$
\begin{eqnarray}\label{instr4}
A_{A'X'}^{YY'ZZ'} =  -\epsilon^{YZ}\frac{(\delta_{X'}^{Z'}\phi^{~Y'}_{A'}+\delta_{X'}^{Y'}\phi_{A'}^{~Z'})}{\phi^2+\omega^2},
\end{eqnarray}
and the four-fermion vertex Eq.(\ref{fieldstrength}) is 
\begin{eqnarray}\label{4fervertex}
	F_{AY'BZ'}^{TT'UU'} =  \frac{4\omega^{2}}{(\phi^2+\omega^2)^2}\epsilon_{AB}\epsilon^{TU}\delta_{(Y'}^{T'}\delta_{Z')}^{U'},
\end{eqnarray}
which is the field strength of an instanton and it satisfies Eq.(\ref{curvature}).

\section{Effect of Anomalies}

We now take up the discussion of anomalies in this model. The Complementary ADHM sigma model of Eqn.(\ref{action4}) is a linear one just like the original model. Neither of these have anomalies. The effective theories of (\ref{seff1}) and  (\ref{seff2}) suffer from chiral anomalies which break spacetime gauge and coordinate invariance unless the gauge field can be embedded in the spin connection of the target space. Anomalies of the original model were discussed in Ref.\cite{Lambert:1995dp}. The anomalies of the effective theory of the present model can be discussed either by the duality transformation or independently following the route in above paper. We shall use a combination of the two.

Since superspace formalism is available only for (0,1) supersymmetry there will be extended supersymmetry anomalies because the (0,4) supersymmetry is not preserved. To cancel these anomalies we have to add finite local counter terms to (\ref{seff2} at all orders of perturbation theory. As a result the spacetime metric and anti-symmetric tensor fields receive corrections at higher orders of $\alpha'$ but the gauge connection does not \cite{Howe:1992tg}.

Other way of looking at this is to notice that even though the action Eq.(\ref{effquantact}) is
classically conformally invariant but when quantized it is not necessarily  ultraviolet finite and the scale invariance is destroyed. According to the  power counting arguments the
off-shell $(0,4)$ supersymmetric models are ultraviolet finite to each order of perturbation
theory \cite{Howe:1987qv}. 
%Sigma model anomalies spoil this argument because only the non-chiral models are ultraviolet finite. The off-shell
%$(0,4)$ supersymmetric theories are one-loop finite but there is a two-loop addition of the
%form $Tr(R^2-F^2)$  that  does not vanish in general. This points to
%non-vanishing $\beta$-functions and we have to take it into consideration while finding the
%conformal fixed point of the renormalization group flow. In models with off-shell
%$(0,4)$ supersymmetry the non-vanishing $\beta$-functions could be canceled by redefining
%the spacetime fields at all orders of $\alpha'$ in such a way as to make sure that supersymmetry
%is conserved in perturbation theory  \cite{Howe:1992tg}. This has been checked up to
%three loops. 
Anomalies in sigma models undermine this argument because only the non-chiral models remain UV finite. Off-shell $(0,4)$ supersymmetric theories maintain one-loop finiteness, but there is an additional two-loop term, $Tr(R^2-F^2)$, which generally doesn't vanish. This implies the presence of non-zero $\beta$-functions, which must be taken into account when seeking the conformal fixed point of the renormalization group flow. In models with off-shell $(0,4)$ supersymmetry, it is possible to cancel these non-zero $\beta$-functions by redefining spacetime fields to ensure the conservation of supersymmetry in perturbation theory, as demonstrated in \cite{Howe:1992tg}. This cancellation has been verified up to three loops. The ultraviolet divergences in the quantization of off-shell
$(0,4)$ models are an artifact of the renormalization scheme which does
not conserve the supersymmetry. The off-shell $(0,4)$ models are ultraviolet finite in a
suitable renormalization scheme. 

The model here has only on-shell $(0,4)$ supersymmetry and these finiteness
arguments do not immediately apply. At least in the $k = k' = 1$ case however the gauge
group $SO(4) \cong SU(2) \times SU(2)$ contains a subgroup $Sp(1) \cong SU(2)$ which allows
three complex structures satisfying the algebra of the quaternions. This gives the target
space of the left-handed fermions with a hyper K\"ahler structure and makes easy an off-shell
formulation utilizing constrained superfields.%\cite{Howe:1988cj}. 
 We may thus anticipate that it is
ultraviolet finite in the similar way as the off-shell models discussed above.

In \cite{Howe:1992tg} Howe and Papadopoulos computed  the field redefinitions  to order  $\alpha'^2$ for $(0,4)$ supersymmetric
sigma models. They found that both  target space metric and antisymmetric tensor field
strength receive corrections to all orders in $\alpha'$. The metric gets the  conformal factor
\begin{equation}\label{conffactor}
\epsilon_{A'B'}\epsilon_{Y'Z'} \rightarrow \left(1-\frac{3}{2}\alpha'f-\frac{3}{13}\alpha'^{2}\Delta f+...\right)\epsilon_{A'B'}\epsilon_{Y'Z'}.
\end{equation} 
%They also studied, up to three-loop order, that these redefinitions cancel the ultraviolet
%divergences that come up when one renormalizes Eq.(\ref{effquantact}) utilizing standard $(0,1)$ superspace
%methods, that do not make sure $(0,4)$ supersymmetry is conserved perturbatively. Moreover, the antisymmetric field strength tensor turns to $H=-\frac{3}{4}\alpha'*df$ such that to cancel
%the gauge anomaly $dH = -\frac{3}{4}\alpha'TrF\land F$. In addition, there  does not exist corrections to the
%instanton gauge field. 
They also investigated, up to the third-order perturbative level, how these adjustments effectively eliminate the ultraviolet divergences that arise during the renormalization of Eq. (\ref{effquantact}) using conventional $(0,1)$ superspace methods. These methods do not guarantee the perturbative conservation of $(0,4)$ supersymmetry. Furthermore, the antisymmetric field strength tensor transforms to $H=-\frac{3}{4}\alpha'*df$ to counterbalance the gauge anomaly $dH = -\frac{3}{4}\alpha'TrF\land F$. Additionally, there are no corrections to the instanton gauge field.

Like the original model the Howe and Papadopoulos expression for
the function $f$ is 
\begin{equation}\label{funcf}
f=-\Delta ln(\phi^2+\omega^2)
\end{equation} 
with $\Delta$  the flat space Laplacian. The corresponding conformal factor of the metric turns out to be
\begin{equation}\label{tsm}
g_{A'Y'B'Z'} = (1+6\alpha'\frac{\phi^2+2\omega^{2}}{(\phi^2+\omega)^2}-18\alpha'^2\frac{\omega^{4}}{(\phi^2+\omega)^4}+...)\epsilon_{A'B'}\epsilon_{Y'Z'}.
\end{equation}

To order $\alpha'$ this agrees with Callan, Harvey and Strominger \cite{Callan:1991dj} expression.
%Remember that the later expression was obtained by solving the first order equations of motion of the ten-dimensional heterotic string (although with $n = 6$ rather than $n = 8$ in their notation). 
Recall that the latter equation was derived by solving the first-order equations of motion for the ten-dimensional heterotic string, albeit with the notation using $n = 6$ instead of $n = 8$.

Thus the flat $R^4$ classical moduli space gets connected to the curved metric in this case too.

\section{Discussion of Small Instantons}
We now take up the case of small instantons. 

Lambert found the order $\alpha'^2$ corrections to the low-energy effective action of the ADHM sigma model. These corrections were in agreement with the earlier results of \cite{Callan:1991ky}. The situation is the same in our case except for the expansion paramenter being
\begin{equation}\label{expansionpara}
\frac{\alpha'}{\Omega(\phi)}=\frac{2\alpha'}{\phi^2+\omega^2}.
\end{equation}
The approximations are valid for all $\phi$ for $\omega^2 \gg \alpha'$ and for $\phi^2 \gg \alpha'$ for small $\omega^2$.

 In Eqn.\ref{tsm} the higher order corrections persist even for $\omega=0$ so that the effective theory is non-trivial just like the original model. In Ref.\cite{Witten:1995gx} Witten had remarked that vanishing of an instanton leads to enhancement of symmetry. Refs.\cite{Galperin:1994qn,Galperin:1995pq} and \cite{Lambert:1995dp} agree upon the idea that the supersymmetry of the sigma model enhances from (0,4) to (4,4) in the infrared limit and the metric must satisfy Laplace equation and be conformally flat.

An alternative view is the following. According to the potentials in (\ref{potential}) the moduli space of the orginal model was given by $\phi=0$ and any $X$. For small instanton $\rho=0$ a new branch of the moduli space opens up for $X=0$ and any $\phi$. This was mysterious. In our model the situation is complementary with the roles of $X$ and $\phi$ reversed. But now we realise that the new branch in small instanton case of the original model is the moduli space of the Complementary branch. Also the new branch in the small instanton case of the Complementary model is the moduli space of the original branch. Hence the mystery was resolved in \cite{Ali:2023csc} since in the small instanton regime the two moduli spaces coalesce while in the finite instanton case these were infinite distance away, that is, disconnected. From this point of view the enhanced symmetry is the $Z_2$ symmetry between the original and the Complementary branches of the moduli space. 

The $(0,4)$ supersymmetry will be enhanced not to $(4,4)$ but from small $(0,4)$ to large $(0,4)$ supersymmetry in case of the complete ADHM sigma model. The enhancement to $(4,4)$ superconformal symmetry has been clarified by Callan, Harvey and Strominger in Ref.\cite{Callan:1991at}. An efficient way of analysing these aspects is to employ the free field realizations of the small, middle and large N=4 superconformal symmetries \cite{Ali:2000zu, Ali:2003aa, Ali:1993sd, Ali:2000we, Hanany:2018hlz} in the conformally invariant infrared limits of these models. The small N=4 superconformal symmetry will be relevant for the conformally flowed infrared limit of  Witten's original and our complementary ADHM sigma models. The large N=4 superconformal symmetry will be relevant for the conformally flowed infrared limit of the complete ADHM sigma model. The middle N=4 superconformal symmetry will be relevant for the Penrose limit of the latter model.

In the small instanton limit of the effective action Eq.(\ref{effquantact}) the field strength (\ref{fieldstrength}) and the $O(\alpha')$ sigma model anomaly vanish. In this limit the metric and the anti-symmetric tensor field are 
\begin{equation}\label{metricg}
g_{\mu\nu} =(1-\frac{3}{2}\alpha'f)\delta_{\mu\nu}~~\text{and}~~ H_{\mu\nu\rho}= -\frac{3}{4}\alpha'\epsilon_{\mu\nu\rho\lambda}\partial^{\lambda}f
\end{equation}
respectively with 
\begin{equation}
    f=-\frac{4}{\phi^2}.
\end{equation}
The metric is of Callan, Harvey and Strominger form but anti-symmetric field is not. In Ref.\cite{Witten:1995zh} and Ref.\cite{Lambert:1995dp} this structure gets connected to the semi-infinte throat geometry of Ref.\cite{Callan:1991at}.

Among the future projects on this topic as well as latest developments we shall mention the following. We have to figure out how to take advantage of the harmonic superspace and projective superspace \cite{Prabhakar:2023kfy} formulations to further clarify the situation. A much more serious issue to deal with concerns the renormalization group flow of both of the models, original and the complementary. The folklore is that these theories should flow to conformal fixed points in the infrared while $N=4$ theories do not flow at all.

In the $AdS_3/CFT_2$ case of the Gauge gravity correspondence \cite{Maldacena:1997re, Witten:1998qj, Gubser:1998bc} present investigations have shed light on the symmetry structure of $AdS_3$ superstrings \cite{Ali:2024amc}. In turn $AdS_3$ superstrings have a close connection with the stringy generalization of the BTZ black hole \cite{Ali:1992mj}.

In Ref.\cite{Papadopoulos:2024uvi} Papdopoulos and Witten gave a direct proof of the fact that in two dimensions scale invariance implies conformal invariance.

\textit{Acknowledgments}: We thank P.P. Abdul Salih for discussions and Prof. D.P. Jatkar and the Harish-Chandra Research Institute, Prayagraj, where part of this work was done, for hospitality. This work was done as part of the Ph.D. work of MI and SRT.

\end{document}